\documentclass[runningheads]{llncs}

\usepackage{dcolumn}
\usepackage{graphicx, caption, tabularx}
\usepackage{multirow}
\usepackage{amsmath}
\usepackage{comment}
\usepackage{subcaption}
\usepackage{float}
\usepackage{xcolor}
\usepackage{longtable}
\usepackage{url}
\usepackage{bm}
\usepackage{multicol}
\usepackage{mathtools, cuted}
\usepackage{stfloats}
\usepackage{caption}
\usepackage{float}
\usepackage{rotating}
\usepackage{lscape}

\usepackage[margin=1in,footskip=0.25in]{geometry}
\newcolumntype{d}[1]{D{.}{.}{#1}}

\begin{document}

%
\title{An Unsupervised Semantic Sentence Ranking Scheme for Text Documents}

%
%
%
\author{Hao Zhang,
Jie Wang\thanks{Corresponding author. E-mail: wang@cs.uml.edu}}
%
%
\institute{Department of Computer Science, University of Massachusetts, Lowell, MA 01854, USA} 
%
\maketitle              %

\begin{abstract}
This paper presents Semantic SentenceRank (SSR), an unsupervised scheme for automatically ranking sentences in a single document according to their relative importance.
In particular, SSR extracts essential words and phrases from a text document,
and uses semantic measures to construct, respectively, 
a semantic phrase graph over phrases and words, and a semantic sentence graph over sentences. It applies two variants of article-structure-biased PageRank to score
phrases and words on the first graph and sentences on the second graph. 
It then combines these scores to generate the final score for each sentence.
Finally, SSR solves a multi-objective
optimization problem for ranking sentences based on their final scores
and topic diversity through semantic subtopic clustering.
An implementation of SSR that runs in quadratic time is presented,
and it outperforms, on the SummBank benchmarks, 
each individual judge's ranking and compares favorably with the combined ranking of all judges. 

\end{abstract}


\begin{multicols}{2}
\section{Introduction}\label{sec:1}

Ranking sentences in a single document according to their relative importance plays a central role in various applications, including summary extraction from a given document (e.g., see  \cite{das2007survey}), structured-overview generation over a large corpus of documents \cite{jww2019multisumm}, and layered reading for fast comprehension of a given document. The last application\footnote{A prototype is available at http://www.dooyeed.com} enables the reader to read a layer of the most important sentences first, then subsequent layers of next important sentences 
until the entire document is read. This application is aimed to facilitate faster reading for understanding.

There are supervised and unsupervised methods for ranking sentences.
Most unsupervised methods use easy-to-compute counting features, such as TF-IDF (term frequency-inverse document frequency)  \cite{neto2000document} and co-occurrences of words \cite{mihalcea2004textrank}. Such approaches are domain and language independent, and are often preferred over other approaches. Not using semantic information in the context, 
these methods may only produce suboptimal sentence ranking. Other unsupervised algorithms  that use semantic information include Semantic Role Labelling \cite{bhartiya2014semantic}, WordNet \cite{bellare2004generic}, and Named Entity Recognition \cite{li2006extractive}. 
These methods, unfortunately, impose a common limitation of language dependence.
In other words, using these algorithms for a given language requires software tools 
to provide the underlying semantic information for the language, which may not be available.
 
Supervised methods require labeled data to train models. Modern supervised methods 
that learn feature representations automatically using a deep neural network model rely on a significant amount of labeled texts (e.g., see \cite{nallapati2017summarunner}). Lacking training data is a major obstacle when developing a supervised sentence-ranking algorithm. The SummBank dataset \cite{radev2003summbank} for evaluating sentence-ranking algorithms, for example, contains only 200 news articles, which is far from sufficient to train a deep neural network model. Other larger datasets 
sufficient to train
neural networks for summarization, 
such as CNN/DailyMail\footnote{Available at https://github.com/abisee/cnn-dailymail}, 
are still unsuitable to train models for ranking sentences, because they only provide a few human-written highlights as a summary for each document.

There are no datasets available at this time for other languages that are 
suitable for training sentence-ranking models.
For example, for the Chinese language, the LSCTS \cite{hu2015lcsts} 
dataset consists of news articles and an average of 1 to 2 sentences written 
by human annotators as a summary for each article; and the NLPCC 2017 dataset\footnote{Available at http://tcci.ccf.org.cn/conference/2017/taskdata.php} also consists of news articles 
with a summary of upto 45 Chinese characters written by human annotators.
These datasets cannot be used to train sentence-ranking models. 
Lacking training data for a particular language has hindered adaption of a good supervised model for one language 
to different languages.

These concerns suggest a direction of investigating unsupervised sentence-ranking algorithms using semantic features that can be computed readily for a given language. 
Word-embedding representations \cite{mikolov2013linguistic} and Word Mover's Distance (WMD) \cite{kusner2015word}, for example, are semantic features of this kind.
 A good word embedding representation provides useful semantic and syntactic information.
Requiring only a large amount of unlabeled texts, it is straightforward to compute word embedding representations for any language using unlabeled out-of-band data such as Wikipedia dumps of the underlying language.
WMD uses word embedding representations to measure semantic distance between two sentences, which can be used to measure their semantic similarity.
 
This paper presents an unsupervised semantic sentence-ranking scheme called Semantic SentenceRank (SSR) using semantic features at the word, phrase, and sentence levels. 
SSR uses phrase and word embedding representations and co-occurrences to construct a semantic phrase-word graph,
scores words and phrases using a variant of article-structure-biased PageRank, 
adjusts scores using Solfplus elevation, and computes a normalized score for each sentence.
SSR then
constructs a semantic sentence graph using WMD, scores sentences using a variant of
article-structure-based PageRank, combines these sentence scores 
to generate the final sentence scores, computes semantic subtopic clustering
of sentences, and ranks sentences by solving a multi-objective 0-1 knapsack problem 
that maximizes the final scores of selected sentences and the diversity of subtopic coverage.

 The major contributions of this paper are as follows:
\begin{enumerate}
	\item {\bf Flexibility}: SSR is an unsupervised scheme using semantic features that can be computed readily for any language.
	
	\item {\bf Efficiency}: SSR runs in quadratic time 
	when using 
	AutoPhrase \cite{shang2018automated} to extract phrases,
	Affinity Propagation \cite{dueck2009affinity} to generate semantic subtopic clusters
	for sentences, and
	a greedy algorithm to approximate the multi-objective 0-1 knapsack problem. 
	
	\item {\bf Accuracy}: Running  on the DUC-02 dataset \cite{DUCdataset}, the aforementioned implementation of SSR  outperforms all previous
algorithms under the ROUGE measures. 
More significantly, on the SummBank dataset \cite{radev2003summbank} , SSR outperforms each individual judge's ranking 
 and compares favorably with the combined sentence ranking of all judges.

\end{enumerate}

The rest of the paper is organized as follows: Section \ref{sec:2} provides a brief overview of related work. SSR is presented in Section \ref{sec:3}. Detail descriptions of the major components of SSR are presented in Sections \ref{sec:4}--\ref{sec:7}. Implementations and evaluation results 
are presented in Section \ref{sec:8}. Conclusions and final remarks are presented in Section \ref{sec:9}.

\section{Related work} \label{sec:2}

Extractive summarization algorithms
that can specify the number of sentences in a summary can be used
to rank sentences,
and vice versa.  

\subsection{Supervised methods} \label{sec:2.1}

Supervised summarization methods can be categorized into two categories: sentence labeling and sentence scoring. 

Supervised sentence-labeling methods assign a binary label to each sentence $S$ to indicate
whether the summary to be produced should include $S$, where the number of sentences with a ``yes" label is determined by the training data. These methods, while being extractive, cannot be used to rank sentences, for they have no control of the number of ``yes" labels to be produced.
Early sentence-labeling algorithms assign labels to each sentence independently
trained on handcrafted features \cite{kupiec1995trainable} such as thematic words and uppercase words. A sequential Hidden Markov Model \cite{conroy2001text} was devised to account for local relations between sentences using three handcrafted features. 
Recently, deep recurrent-neural-network models \cite{cheng2016neural,nallapati2017summarunner,narayan2017neural} were used to derive a meaningful representation of a document and carry out sequence labeling 
based on labels of previous sentences. 
Trained with cross-entropy loss, however, these models were 
redundancy-prone and tend to generate 
verbose summaries 
 \cite{narayan2018ranking}. 
Supervised sentence-labeling methods also have the following two downsides. First, labeled datasets that are large enough to train a model may be hard to come by. Second, labeled datasets required to train a model that exist for one language may not exist for a different language, making it difficult to adapt a model to different languages.

Supervised sentence-scoring methods depend on
the underlying similarity measure of a sentence in a document to 
a benchmark summary.
For example,
CNN-W2V \cite{zhang2016extractive} is a model that computes a ROUGE score for each sentence in a document using the corresponding summaries in a labeled data as references,
and uses such scores as the ground truth to train a convolutional-neural-network model to score sentences independently of input documents. This means that the same sentence appearing in 
different documents always has the same score. This is problematic, for the same sentence
in different documents is unlikely to be equally important.
A more reasonable approach is to score sentences via global optimization by taking  previously scored sentences and their scores into consideration.
Refresh \cite{narayan2018ranking}, for example, is a recent model in this direction.
It scores a sentence using previously scored sentences
and the summaries of the underlying document in a labeled dataset, where
sentence scores are used as a reward function in the model. 

No matter what the underlying method is, it is necessary to have
a large labeled dataset to train a supervised neural network model.
 This necessity remains a major obstacle, for such a dataset
may not exist for a given language.

\subsection{Unsupervised methods}

Unsupervised summarization methods exploit relations between words, as related words ``promote'' each other. For example, TextRank \cite{mihalcea2004textrank} and LexRank \cite{erkan2004lexrank} each models a document as a sentence graph based
on word relations, but they use only syntactic features.
PageRank \cite{PageRank} is used to score words. 

Other methods incorporate additional information for achieving a higher accuracy.
UniformLink \cite{wan2010exploiting}, for example, constructs a sentence graph on a set of similar documents, where a sentence is scored based on both of the in-document score and cross-document score. URank \cite{wan2010towards}, on the other hand,  uses a unified graph-based framework to study both single-document and multi-document summarization. 

The quality of a summary may be improved using max-margin methods \cite{li2009enhancing} or integer-linear programming (ILP) \cite{parveen2015integrating, parveen2015topical}. Among the previous algorithms, $CP_3$ \cite{parveen2016generating} offers the highest ROUGE-1, ROUGE-2, and ROUGE-SU4 scores over DUC-02. It uses a bipartite graph to represent a document, and a different algorithm, Hyperlink-Induced Topic Search (HITS) \cite{kleinberg1998authoritative}, is used to score sentences. $CP_3$ treats the summarization problem as an ILP problem, which maximizes the sentence importance, non-redundancy, and coherence simultaneously. However, since solving ILP is NP-hard, obtaining an exact solution to an ILP problem is intractable. 

It is worth noting that a recent unsupervised method \cite{Isonuma2019}, while not related to the problem studied in this paper, may provide new ideas. On a collection of consumer reviews on a particular product, the method generates an abstractive sentence as the main review point, and ranks sentences by the number of descendants in a discourse tree
rooted on the main review point.
It would be interesting to investigate if this method can be modified to rank sentences of a given document according to their relative importance. 

Early unsupervised methods have two common downsides: 
\begin{enumerate}
\item They don't promote diversity. This is because the importance of sentences are based only on sentence scores, and so sentences of high scores representing the same subtopic may all be included in a summary, leaving no room to include sentences with lower scores but with different subtopics.
\item These methods do not used semantic features.
\end{enumerate}

\section{Semantic SentenceRank} \label{sec:3}

To overcome the downsides of the existing methods (supervised or unsupervised), a better sentence-ranking algorithm
should incorporate semantic features and topic diversity, and it should be unsupervised.

Let $D$ denote a document consisting of $n$ sentences indexed as $S_1, S_2, \ldots, S_n$ in the order they appear, each with a length $l_i$, along with a maximum length capacity $L$, where $l_i$ is the number of characters contained in $S_i$. Let $F_{s}(S_i)$ and $F_{d}(D)$ denote a semantic sentence scoring function and a diversity coverage measure, respectively. 
Then the semantic sentence-ranking problem is modeled as follows:
\begin{align*}
\text{maximize} &\sum_{i=1}^n F_s(S_i)x_i \mbox{ and } F_d(D), \\
\text{subject to} & \sum\limits_{i=1}^n l_ix_i \leq L \mbox{ and }x_i \in \{0, 1\}.
\end{align*}
where $x_i$ is a 0-1 variable such that $x_i = 1$ if sentence $S_i$ is selected, and 0 otherwise.
By setting $L$ appropriately from small to large, one can obtain from solving the
optimization problem the first sentence, then the second, then the third, and so on until all sentences are ranked. Unfortunately, this problem is NP-hard and so an approximation algorithm is needed. 

{SSR computes $F_s$ by combining salience scores 
at three levels: words, phrases, and sentences. 
In particular, it first constructs a semantic phrase-word graph (SPG) on phrases and words,
and a semantic sentence graph (SSG) on sentences.
It then computes $F_d$ using semantic subtopic clustering.
Finally, SSR uses an
\parfillskip=0pt\par}
\end{multicols}

\begin{figure}[H]
\centering
\begin{tabular}{cc}
\includegraphics[width=0.49\textwidth]{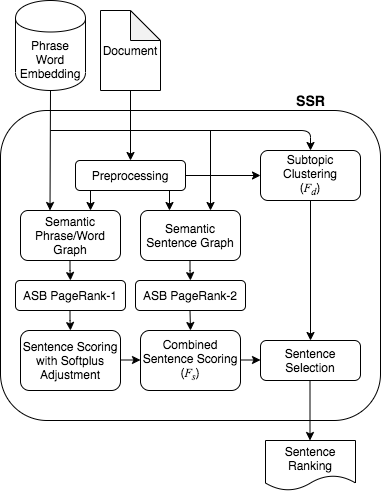} ~~&~~
\includegraphics[width=0.45\textwidth]{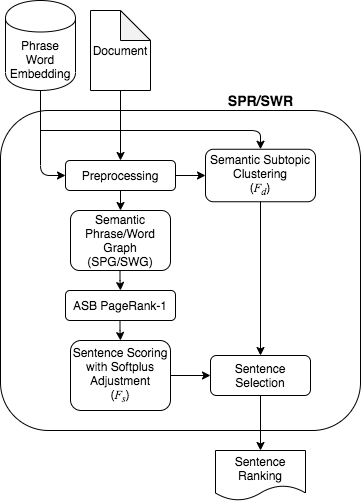} \\
\\
(a) Data flow diagram for SSR & (b) Data flow diagram for SPR/SWR\\
\end{tabular}
\caption{Major components of (a) SSR and (b) SPR/SWR}
\label{fig:SSR}
\end{figure}

\begin{multicols}{2}
\noindent 
approximation algorithm based on $F_s$ and $F_d$
to rank sentences.
Fig. \ref{fig:SSR}(a)
depicts the data flow diagram for the major components of SSR.

\subsection{Sub-models}

SSR contains two sub-models: one at the word level known as Semantic WordRank (SWR) \cite{zhang-wang2018}, 
and one 
at the phrase-word level referred to as Semantic PhraseRank (SPR).
In other words, SPR is SSR excluding semantic sentence graph 
and ABS-biased PageRank-2. Both SWR and SPR follow the same data follow diagram (see Fig. \ref{fig:SSR}(b)),
except that SWR does not consider phrase-level similarities.  Both are faster than SSR, and
perform well on selecting top-ranked sentences, which is sufficient for certain applications.

\subsection{Phrase and word embedding}

A searchable dataset of phrase and word embedding representations is calculated independently of
SSR. Such an embedding dataset may be available for free download for some languages. If unavailable, it
is straightforward to compute word and phrase embedding over an unlabeled Wikipedia dump using a standard method. To extract phrases, a linear-time unsupervised algorithm such as AutoPhrase \cite{shang2018automated} may be used.
Recalculations of phrase and word embedding may be carried out once in a while on a larger Wikipedia dump.

\subsection{Preprocessing}
The preprocessing component computes, on a given document $D$, 
the phrases contained in $D$ using a phrase extractor, and
the set of essential words
contained in $D$ that, excluding these phrases, pass a part-of-speech (POS) filter, a stop-word filter, and a stemmer for reducing inflected words to the word stem. It removes all non-essential words. In what follows, unless otherwise stated, when words are mentioned, they are essential words.

Descriptions of the remaining components of SSR
are presented in Sections \ref{sec:4}--\ref{sec:7}. 
 
\section{Semantic graph representations} \label{sec:4}

\subsection{Semantic word graph}

In addition to considering co-occurrence between words as in TextRank \cite{mihalcea2004textrank}, 
the semantic word graph (SWG) for the underlying 
document
adds 
embedding similarity of words 
to enhance connectivity of the graph. 
Adding semantic similarity is vital for processing analytic languages, such as Chinese, which seldom use inflections. When stemming is not applicable, semantic similarity serves as an alternative to represent the relations between words with similar meanings, allowing them to share the importance when computing PageRank scores for these nodes.

Let  
$G = (V, E)$ be a weighted graph of words in document 
$D$. 
Two words in $V$ are connected if either they co-occur within a window of $\Delta_{\text{SWG}}$ successive words in the document (e.g, $\Delta_{\text{SWG}} = 2$), or the cosine similarity of their embedding representations exceeds a threshold value $\delta_{\text{SWG}}$ (e.g. $\delta_{\text{SWG}} = 0.6$). 

Let $u$ and $v$ be two adjacent nodes. 
For each edge $(u,v)$, if only one type of connection exists, then treat the weight of the other type 0.
Assign the co-occurrence count of $u$ and $v$ as the initial weight to the co-occurrence connection and the cosine similarity value as the initial weight to the 
semantic connection. Normalize the initial weights of co-occurrence connections; namely, 
divide the initial co-occurrence weight by the total initial co-occurrence weight.
Normalize the initial weights of semantic connections; namely, divide the initial semantic weight by the total
initial semantic weight. Let $w_c(u,v)$ and $w_s(u,v)$ denote, respectively,
the normalized weight for the co-occurrence connection and the semantic connection of
$u$ and $v$. Finally, assign $$w(u,v) = w_c(u,v)+w_s(u,v)$$ as the weight to
the edge $(u,v)$.

\subsection{Semantic phrase graph}

In a SWG, 
words in a phrase (e.g., names, scientific terms, and general entity names) would have high co-occurrence counts 
if the phrase appears multiple times in the document. The meaning of a word
inside a phrase may be different from that outside the phrase.
For example, in the sentence ``There is an apple on top of her Apple computer",
the word ``Apple" appears outside and inside the phrase of ``Apple computer", 
which has different meanings.
Thus, a high-quality phrase extractor is desired when building a phrase graph. 

Given a document $D$, SSR applies a phrase extractor 
to segment phrases in $D$. Let $P$ denote the set of phrases 
and $W$ the set of words in $D$ after phrases are removed. 
If a phrase $p  \in P$ of the given document does not appear in 
the database of phrases,
then remove $p$ from $P$ and add the words $w\in p$ to $W$ (Note that the probability
of this to happen is small if the construction of the database of phrase embedding uses
the same phrase extractor.

A semantic phrase-word graph (SPG) is
a weighted 
graph $(V,E)$ with $V  = P\cup W$ such that
two nodes are connected if either they co-occur in a small sliding window of $\Delta_{\text{SPG}}$ consecutive
words and phrases 
or the cosine similarity of their embedding representations is greater than a threshold value 
$\delta_{\text{SPG}}$. 

\subsection{Semantic sentence graph}

A semantic sentence graph (SSG) of a document $D$ is a weighted graph with sentences in $D$ being its nodes, where two sentences $S_i$ and $S_j$ are connected if either they contain
a common word or phrase, or the WMD of $S_i$ and $S_j$
is below a certain value, which may be determined by how large a percentage of sentences
should be connected. 
The weight of an edge $(S_i,S_j)$ is determined as follows:
\begin{enumerate}
\item Let $p_{ij}$ denote the number of phrases contained in both $S_i$ and $S_j$. 
After removing common phrases, let $v_{ij}$ denote the number of words contained in both
$S_i$ and $S_j$. Let $|S_i|$ and $|S_j|$ denote, respectively, the number of words contained in
$S_i$ and $S_j$. Let
\begin{equation*}
w_c(i,j) = \frac{p_{ij} + v_{ij}}{\log_{10} |S_i| + \log_{10} |S_j|}.
\end{equation*}

\item  Define a different similarity measure of $S_i$ and $S_j$ by
\begin{equation}
\hspace*{-2mm}\text{sim}_P(S_i,S_j) = \frac{1}{1+\text{WMD}(S_i,S_j)}.\label{eq:5}
\end{equation}
Sort the similarity scores given by Eq. (\ref{eq:5}) in descending order. Select a $\Gamma\%$
of the edges with corresponding similarity scores being the top $\Gamma\%$ of the similarity scores (e.g., $\Gamma = 30$).
Add these semantic edges to the graph with weights being the corresponding similarity scores.

\item Normalize the co-occurrence edge weight $w_c(i,j)$; namely, divide $w_c(i,j)$ by
$\sum_{i\not= j} w_c(i,j)$. Normalize the semantic edge weight; namely, divide 
the similarity given by Eq. (\ref{eq:5}) by the total similarity weight of all semantic edges.
Sum up the two normalized weights to be the final edge weight $w_{ij}$.
\end{enumerate}

\section{Sentence scoring} \label{sec:5}
\subsection{Article-structure-biased PageRank-1}

Article structures define how information is presented. For example, the typical structure of news articles is an inverted pyramid \cite{po2003news}, where critical information is presented at the beginning, followed by additional information with less important details. In academic writing, the structure of an article would look like an hourglass\footnote{See, for example,
discussions in
\url{https://www.unbc.ca/sites/default/files/assets/academic_success_centre/writing_support/hourglass.pdf}.}, which includes an additional conclusion piece at the end of the article.
Thus, sentence locations in an article according to the underlying structure also plays a role in ranking sentences. SWR uses a position-biased PageRank algorithm \cite{florescu2017position}.

Directly applying PageRank, one can compute a score $W(v_i)$ of a node $v_i \in G$
by iterating the following equation until converging:
\begin{align}
\hspace*{-2mm}
W(v_i) =&  d \bigg(\sum_{v_j \in Adj(v_i)} \frac{w_{ji}}{\sum\limits_{v_k \in Adj(v_j)} w_{jk}} W(v_j)\bigg)+\nonumber
 \\
 &+ (1-d),\label{eq:1}
\end{align}
where $Adj(v_i)$ denotes the set of nodes adjacent to $v_i$, $d \in (0,1)$ is a damping factor, and $w_{ji}$ is the weight of the edge between node $j$ and node $i$.
The value of $d$ is set to 0.85 as in the original PageRank paper \cite{brin2012reprint}
and the TextRank paper \cite{mihalcea2004textrank}.
The intuition behind this equation is that the importance of a node $v_i$ is affected by the scores of its adjacent nodes and the probability of $1-d$ for jumping from a random node to node $v_i$.

Eq. (\ref{eq:1})  is an unbiased PageRank, where each word is assumed equally likely to start from. In article-structure-biased (ASB) PageRank, each word $v_i$ is biased with a probability $P(v_i)$ according to the underlying article structure. For example, in the inverted pyramid structure, a higher probability 
is assigned to a word that appears closer to the beginning of the article.

Rank the importance of sentence locations from the most important to the least important based on the underlying article structure (Note: This is not the sentence ranking to be computed). 
Let $\text{LS}_i(w)$ denote the location score of $w \in S_i$, where $S_i$ is the $i$-th sentence.

The probability for node $v_i$ can now be computed by
\begin{align*}
P(v_i) = \frac{\sum_{k:v_i \in S_k} {\text{LS}_k(v_i)}}{\sum_{j,k:v_j \in S_k} {\text{LS}_k(v_j)}}.
\end{align*}
Note that the above computation is at the sentence level, which can be easily adapted to the word level by ranking words instead of sentences.

The ASB PageRank score $W'(v_i)$ for node $v_i$ is computed as follows:
\begin{align}
\hspace*{-4mm}
	W'(v_i) = & d\bigg(\sum_{v_j \in Adj(v_i)} \frac{w_{ji}}{\sum_{v_k \in Adj(v_j)} w_{jk}} W'(v_j) \bigg)+\nonumber
\\ &+(1 - d) P(v_i).\label{eq:2}
\end{align}
Computation starts with an arbitrary initial value for each node, and iterates the computation
of Eq. (\ref{eq:2}) until it converges. 

$W'(v_i)$, referred to as \textsl{salient score}, represents its importance relative to the other words in the document.  

\subsection{Softplus adjustment}

Let $S$ be a sentence. To score $S$, one may simply sum up the salient score of each word contained in $S$ and normalize it by $|S|$ (the number of essential words contained in $S$).
Normalization ensures that longer sentences and shorter sentences are comparable (otherwise, larger scores may have larger scores just because they have more words).
Namely, let 
$$\text{sal}(S) = \frac{1}{|S|}\sum_{v_i \in S} W'(v_i).$$
This way of scoring, however, has a drawback. To see this, 
suppose that $S_1$ and $S_2$ are two sentences with similar scores under
this method, and contain about the same number of words.
If the distribution of word scores for words contained in $S_1$ 
follows the Pareto Principle, namely,
a few words have very high scores and the rest have very low scores close to 0,
while $S_2$ has roughly a uniform word score distribution, where the high scores of a few words in $S_1$
are much larger than the (almost uniform) scores of words in $S_2$, then
the few words in $S_1$ with very high scores would make $S_1$ appear more important than $S_2$. Using direct summation of salient word scores, 
it is possible to end up with the opposite outcome. 

Using the Softplus function $sp(x) = \ln (1+e^x)$ helps overcome this drawback \cite{SZJW2017}. Commonly used as an activation function in neural networks, $sp(x)$ offers a significant elevation of $x$ when $x$ is a small positive number. If $x$ is large,  then $sp(x) \approx x$.

Apply the Softplus function to each word, and sum up the elevated values to 
be the salient score of $S$, denoted by $\text{sal}_{sp}(S)$. Namely, 
\begin{equation}
\text{sal}_{sp}(S) = \frac{1}{|S|}\sum_{v_i \in S} 
\ln(1 + e^{W'(v_i)}). \label{eq:3}
\end{equation}

To illustrate this using a numerical example, assume that $S_1$ and $S_2$ each consists of 5 words, with
original scores ($W'$) and Softplus scores ($sp' = sp\circ W'$) given in the following table (Table \ref{example}):

\noindent
\begin{center}
\captionof{table}{Numerical examples with $W'$ and $sp'$ scores}
\label{example} 
\begin{tabular}{l||c|c|c|c|c||c}
\hline
$S_1$ & $v_{11}$ & $v_{12}$ & $v_{13}$ & $v_{14}$ & $v_{15}$&   sal \\
\hline
$W'$ & 2.6 & 2.2 & 2.1 & 0.3 & 0.2 & 1.48\\
$sp'$ & 2.67 & 2.31 & 2.22 & 0.85& 0.80 & \bf 1.768\\
\hline
$S_2$ & $v_{21}$ & $v_{22}$ & $v_{23}$ & $v_{24}$ & $v_{25}$ & \\
\hline
$W'$ & 1.6 & 1.5 & 1.5 & 1.5 & 1.4  & \bf 1.5  \\
$sp'$ & 1.78& 1.70& 1.70& 1.70& 1.62&  1.702 \\
\hline
\end{tabular}
\end{center}
Sentence $S_1$ is more important than $S_2$
because it contains three words of much higher $W'$-scores than those of $S_2$.
However, $\text{sal}(S_1) = 1.48 < \text{sal}(S_2) = 1.5$ and so $S_2$ will be
selected.
After using Softplus, $\text{sal}_{sp}(S_1) = 1.768 > \text{sal}_{sp} (S_2) = 1.702$,
and so
$S_1$ is selected as it should be. Experiments (see Section \ref{sec:feature}) indicate that
using the Softplus elevation does improve ranking accuracy in practice.

\subsection{Article-structure-biased PageRank-2}

For the semantic sentence graph, SSR uses a modified ASB PageRank algorithm to score sentences, as ASB PageRank-1 suitable for words may not be suitable for sentences. To see this, assume that a document has the inverted pyramid structure, then
using the reciprocal of the location index of a sentence as its location score will
result in putting too much weight on the first few sentences and too little weight on the subsequent sentences. Clearly, this Pareto phenomenon is not practical. Instead,
let $\text{LS}(S_i)$ denote
the location score of sentence $S_i$ (recall that the subscript $i$ is the location index of
the sentence).
Normalize $\text{LS}(S_i)$ to generate $P(S_i)$ for the modified ASB PageRank algorithm to score $S_i$ similar to Eq. (\ref{eq:2}) as follows:
\begin{align*}
\hspace*{-8mm} W'(S_i) = & d \bigg(\sum_{S_j \in Adj(S_i)} \frac{w_{ji}}{\sum_{S_k \in Adj(S_j)} w_{jk}} W'(S_j)\bigg) +
\\ &+(1 - d) P(S_i).
\end{align*}

\subsection{Combined sentence scoring}

The sentence scoring function $F_s$ is defined as follows:
For any given sentence $S$ contained in $D$, 
\begin{eqnarray}
F_s(S) = \frac{1}{2}\left(\text{sal}_{sp}(S) + W'(S)\right).\label{eq:Fs}
\end{eqnarray}

\subsection{A remark on other node centrality measures}

PageRank is a node-centrality measure. Other centrality measures may also be used to score nodes on a given SWG, SPG, and SSG, including
degree centrality, closeness centrality, eigenvector centrality, and diffusion centrality, among others (see, e.g., \cite{bloch2016centrality}). However, adding article-structure-biased information of a node using these methods is not as natural as using PageRank.

\section{Semantic subtopic clustering} \label{sec:6}

Selecting a sentence based only on sentence scores would result in a poor diversity
of topic coverage, as multiple sentences of the same subtopic could have higher scores than
sentences of different subtopics. 
To avoid this drawback, a sentence subtopic clustering method is needed. 

Clustering algorithms depend on a chosen similarity measure for the underlying objects
to be clustered. Clustering may be carried out based on thematic similarity measures or semantic similarity measures. Thematic clustering groups sentences of the same context into the same cluster. For example, under thematic similarity measures, sentences that contain the following words may be
grouped into the same cluster: frog, pond, green, or tree \cite{Tinkham1997}. 
TextTiling \cite{hearst:97}, for example, is a thematic clustering algorithm.
First used in text summarization \cite{SZJW2017},
TextTiling groups several consecutive paragraphs into the same cluster by finding
thematic shifts between consecutive paragraphs. Unfortunately, it often fails to generate multiple clusters on short articles or when there are no clear thematic shifts between consecutive paragraphs. 

Clustering methods based on TF-IDF over the BOW (bag-of-words) representations are in general unsuitable for measuring document distances or similarities  due to
frequent near-orthogonality \cite{kusner2015word, greene2006practical}.

Semantic clustering, on the other hand,  
groups sentences that have similar meanings or convey similar information
into the same cluster. For example,  under semantic similarity measures, sentences about
eyes, noses, and ears may be group into the same cluster.
Semantic clustering is based on semantic measures between sentences.
WMD, in particular, can be used to define semantic measures.

Efficiency, accuracy, and easy implementation are
criteria to choose a clustering algorithm. 
When choosing a clustering algorithm to implement SSR, 
it is imperative to choose one that can also be easily modified to use semantic measures. Moreover, the complexity of the algorithm should not exceed quadratic time, as higher complexity may cause interactive applications (such as the layered-reading tool at http://www.dooyeed.com) unacceptable in practice.  Note that pairwise comparisons of objects alone in a clustering algorithm require a quadratic-time lower bound. 
 
Spectral clustering \cite{von2007tutorial} and affinity propagation \cite{dueck2009affinity} are
both based on k-means using Euclidean distance as the underlying similarity measure, 
which can easily be replaced with a semantic measure such as WMD, and they can be
carried out in quadratic time.  Any
clustering method that possesses these properties may also serve as candidates. Other context-aware similarity measures such as the one described in \cite{Besbes2016} may be explored. 
The topic diversity function $F_d$ can be represented using such a semantic subtopic clustering algorithms.

\subsection{Semantic spectral clustering} \label{sec:spectral}

Spectral clustering \cite{von2007tutorial} uses eigenvalues of a similarity matrix (aka. affinity matrix) to reduce dimension before clustering a given set of data points into $K$ clusters, where
$K$ is a preset positive integer. Spectral clustering can
handle data points that do not satisfy convexity.

Treat each sentence as a data point.
Let $\text{WMD}(S_i,S_j)$ denote the Word Mover's Distance between two sentences $S_i$ and $S_j$. 
The similarity matrix is an $n\times n$ matrix, where $n$ is the total number of sentences in a document, and the entry $s_{ij}$ of the matrix corresponds to a similarity measure 
between two sentences $S_i$ and $S_j$ defined in Eq. (\ref{eq:sim1}). Under the WMD metric, a smaller value between two sentences means that they  are more similar, while a larger value means 
that they are less similar. This
can be transformed to a similarity metric using the RBF kernel as follows:
\begin{equation}
\text{sim}_G(S_i, S_j) = e^{-\gamma \cdot \text{WMD}(S_i, S_j)^2}, \label{eq:sim1}
\end{equation}
where $\gamma$ may be set to 1.
It then uses k-means to generate clusters over eigenvectors corresponding to the $K$ smallest eigenvalues.

The number of clusters $K$ is related to $n$. Empirical studies suggest that 30\% of the original text size would be the best size for a summary to contain almost all significant points contained in a single document. In other words, extracting about $0.3n$ sentences
appropriately would cover almost all key points in the original document.
On the other hand, to avoid having too many clusters that could deteriorate performance, it is necessary to set an upper bound $C$. For typical news articles, for example, an upper bound $C= 8$ would be appropriate.
Thus, let
$$K = \min\{\lfloor0.3n\rfloor, C\}.$$

Let $n_i$ denote the number of non-repeated essential words contained in sentence $S_i$,
which is bounded above by a constant $M$ (e.g., $M = 30$. If in a rare occasion $n_i$
is longer than this bound, one can split the sentence into natural clauses). 
The time complexity of computing $\text{WMD}(S_i,S_j)$ is $O(M^3) = O(1)$.
Hence, $\text{sim}_G(S_i,S_j)$ defined in Eq. (\ref{eq:sim1}) and $\text{sim}_P(S_i,S_j)$ defined in Eq. (\ref{eq:5}) can both be computed
in $O(1)$ time. 

Thus, computing
the similarity matrix incurs $O(n^2)$ time.
Using the implicitly restricted Lanczos method \cite{LS}, finding the $K$ largest eigenvalues and the corresponding eigenvectors over an $n\times n$ symmetric real matrix can be done in $O(Kn^2)$ time.
There are a number of heuristic algorithms to approximate k-means  that run in $O(Kn^2)$ time \cite{kmeans2019}. Since $K$ is set to be less than a constant $C$,  
semantic spectral clustering can be carried out in quadratic time.

\subsection{Semantic affinity propagation} \label{sec:ap}

Recall that spectral clustering 
must fix a number of clusters before clustering. This could be problematic in practice. Affinity propagation (AP) clustering \cite{dueck2009affinity} overcomes this problem. It is an exemplar-based clustering algorithm such as k-means  \cite{hartigan1979algorithm} and k-medoids \cite{kaufman2009finding} except that AP does not need to preset the number of clusters.

Let $S_1,S_2,\ldots,S_n$ be the sentences to be clustered under the similarity measure
of sim$_P(S_i,S_j)$ defined in Eq. (\ref{eq:5}), which runs in constant time.
Each sentence $S_i$ is a potential
exemplar and let $\text{sim}_P (S_i,S_i) = m$, where $m$ is the median of 
$\text{sim}_P(S_i,S_j)$ 
for all $i, j\in \{1,2,\ldots,n\}$ with $i \not= j$. AP proceeds by updating two $n\times n$ matrices 
$\bm{R} = (r_{ij})$ (the responsibility matrix) and $\bm{A} = (a_{ij})$ (the availability matrix)
as follows until they converge for all $i$ and $j$:
\begin{enumerate}
\item Initially, set $r_{ij} \leftarrow 0$ and $a_{ij} \leftarrow 0$.
\item Set $r_{ij} \leftarrow \text{sim}_P(S_i,S_j) - b_{ij}$, where 
$$b_{ij} = \max_{j' \not= j} \{\text{sim}_P(S_i,S_{j'}) + a(i,j')\}.$$
\item If $i \not= j$, then set 
$$a_{ij} \leftarrow \min\{0, r_{jj}\} + \sum_{i' \not\in\{i,j\}} \max\{0,r_{i'j}\}.$$
Otherwise, set
$$a_{ij} \leftarrow \sum_{i' \not= j} \max\{0,r_{i'j}\}.$$

\end{enumerate}
If $r_{ii} + a_{ii} > 0$, then $S_i$ is selected as an exemplar and $S_j$ belongs to the
cluster of $S_i$ if $S_j$ has the largest similarity with $S_i$ among all other
exemplars.
Semantic AP runs in $O(n^2)$ time.

\section{Sentence selection and ranking} \label{sec:7}

An approximation algorithm is needed to cope with the NP-hardness of the multi-objective 0-1 knapsack problem. There are various techniques for tackling
multi-objective optimization problems such as those described in \cite{Miettinen1999,Branke2008,chen2015, rostami2017progressive, liu2019convergence, zhang2014optimization}. Other recent optimization techniques on solving integrated computer-aided engineering problems include 
\cite{Rostami2016,Pan2017,Liu2018,Yan2019}.
A good approximation algorithm should balance between accuracy and efficiency. 
The following greedy approximation in a round-robin style is used in the current implementation of SSR.

\subsection*{Round-robin selection}
Each sentence $S_i$ is now associated with four values: (1) sentence index $i$, (2) salient score $F_s(S_i)$ computed by Eq. (\ref{eq:Fs}), (3) sentence length $l_i$, and (4) cluster index $j$ of the cluster $S_i$ belongs to. Select sentences greedily in a round robin fashion and rank them as follows:

\begin{enumerate}
\item Let ${\cal S}$ denote the set of selected sentences. Initially, ${\cal S} \leftarrow \emptyset$.
\item For each sentence $S_i$, compute the value per unit length to obtain a unit score $s'_i = s_i/l_i$.
\item For each cluster $c_j$, sort the sentences contained in it in descending order according to their unit scores.
\item \label{s:4} 
While there are still sentences that have not been selected, do the followings:
\begin{enumerate}
\item Sort the remaining clusters in descending order 
according to the highest unit score contained in a cluster. 
For example, if the highest unit score in cluster $c_i$ is smaller than the highest unit score in cluster $c_j$, then $c_j$ comes before $c_i$ in the sorted clusters.
\item \label{s:5}
Select the sentence from the remaining sentences with the highest unit score, one from each cluster in the order of sorted clusters, and add it to $\cal S$. That is, 
$${\cal S} \leftarrow {\cal S} \cup \{S_{i_1}, S_{i_2}, \ldots, S_{i_k}\},$$ where $S_{ij}$
are the selected sentences and $k$ is 
the number of remaining clusters that are nonempty. 

\item \label{s:6} Remove the selected sentences from their corresponding clusters.
\end{enumerate}

\item Rank sentences according to the order they are selected.
\end{enumerate}

\section{Implementations and evaluations} 
\label{sec:8}

\subsection{Embedding database and parameters}

The word-embedding database uses a pre-computed word embedding representations with subword information \cite{bojanowski2016enriching}, which can handle out-of-vocabulary words and generate better word embedding for rare words. 

The phrase-embedding database uses 
AutoPhrase \cite{shang2018automated} on an English Wikipedia dump and the CNN/DailyMail dataset to extract phrases and words. fastText \cite{bojanowski2016enriching} is then used to compute embedding representations for these phrases and words with respect to the same datasets. AutoPhrase is an unsupervised phrase extractor that supports any language as long as a general knowledge base and a pre-trained POS-tagger (recall that POS stands for 
part-of-speech) in that language are
available. Wikipedia is typically used as a general knowledge base and pre-trained POS-taggers are widely available for different languages.

The sliding-window size for computing co-occurrence of words for SWG is set to $\Delta_\text{SWG} = 2$,
and the sliding-window size for computing co-occurrence of words for SPG is set to $\Delta_\text{SPG} = 3$.
As noted in constructing TextRank word graph on co-occurence \cite{mihalcea2004textrank}:
``A larger window does not seem to help---on the contrary, the larger the window, the lower the precision, probably explained by the fact that a relation between words that are further apart is not strong enough to define a connection in the text graph."
Setting a window size of 2 to capture co-occurrence for words was recommended.
Because most phrases consist of two words, the window size to capture
co-occurrence of phrases and words should be just larger than 2, hence
setting the window-size to 3 for SPG is reasonable. Note that setting window sizes 
slightly larger may slightly degrade the precision.

A  cosine similarity value of word or phrase embedding that is larger than 0.6 is deemed
sufficient to indicate two words are semantically similar, which ensures that the
underlying semantic graph has sufficient connectivity, but not too dense. 
Because there are more words than
phrases, setting the threshold value of semantic similarity for words to 0.65 and
for phrases to be 0.6 is reasonable. Note that setting the threshold values slightly larger will only slightly affect the precision. 
 
For the semantic sentence graph, 
setting the percentage $\Gamma\% = 30\%$ 
provides sufficient connectivity. 

\subsection{Implementations and complexity analysis}

The implementation of SPG uses AutoPhrase to extract phrases. 
A straightforward table lookup
of the embedding database provides the embedding representations of words and phrases
needed to construct SPG. 

Since the datasets to be used to evaluate sentence-ranking algorithms are
news articles and news articles in general have the inverted pyramid structure,
to implement ABS-biased PageRank-1,  the following
location score for word $w$ in the $i$-th sentence $S_i$ is used:
$$\text{LS}(w|S_i) = \frac{1}{i}.$$
Likewise, to
implement ABS-biased PageRank-2, the following location score for sentence $S_i$
is used:
$$\text{LS}(S_i) = \frac{1}{\log_{10} (1+i)}.$$
Note that location scores may be defined using different functions. For example,
the structure of an academic research paper may in general have the hourglass 
structure\footnote{\url{https://www.unbc.ca/sites/default/files/assets/academic_success_centre/writing_support/hourglass.pdf}.}.

Finally, semantic special clustering is used to implement SWR and SPR, while
semantic affinity propagation is used to implement SSR.

Under the said implementations, SWR, SPR, and SSR on a given document $D$ all run in $O(|D|^2)$ time, where
$|D|$ is the number of essential words contained in $D$.
This can be shown as follows:
\begin{enumerate}
\item The preprocessing of extracting phrases using AutoPhrase can be done in $O(|D|)$ time,
so does extracting words.

\item The embedding database may be implemented as a dictionary, and so looking-up a word $w$ 
takes $O(1)$ time. Looking-up a phrase is similar. Thus, retrieving embedding representations of all $m$ different words and phrases contained in $D$
takes $O(m)$ time.
Constructing a SPG takes $O(m^2)$ time (constructing a SWG is similar), and constructing SSG takes $O(n^2)$ time,
where $n$ is the number of sentences. Both $m$ and $n$ are less than $|D|$.

\item The running time of both ABS-biased PageRank algorithms is $O(\ell|D|^2)$, where
$\ell$ is the number of iterations, which tends to be small in practice, and so can be
considered as a constant. (One may also fix a reasonable number of iterations
and use whatever the values returned at the end of the last iteration. This is sufficient in practice.)

\item The Softplus adjustment can be done in $O(|D|)$ time.

\item Semantic spectral clustering runs in $O(|D|^2)$ time (see Section \ref{sec:spectral}).

\item Semantic affinity propagation runs in $O(n^2) < O(|D|^2)$ time (see Section \ref{sec:ap}). 
\end{enumerate}

To shortern the running time, a linear-time relaxed version of Word Mover's Distance \cite{atasu2017linear} is used in the experiments and evaluations.

\subsection{Datasets for evaluation}

Most datasets for evaluating summarization algorithms
consist of one or more human-written summaries for each text document. These summaries either have a fixed number of words or a fixed number of sentences. For example, each summary in DUC-02 consists of about 100 words or less. Some datasets, such as 
CNN/Daily Mail, may only contain summaries of one to three human-written sentences.

To evaluate a sentence-ranking algorithm using such datasets, one may use an appropriate
number of sentences of the highest rank it produces to match the size of the underlying summary and
compare their ROUGE scores. 
This approach, however, 
cannot be used to establish the accuracy of sentence ranking for all sentences in a document.

It is customary to use the DUC-02 benchmarks to evaluate the effect of summarization algorithms, including extractive summarization, even though DUC-02 benchmarks are abstractive summaries. In particular, DUC-02 contains a total of 567 news articles with an average of 25 sentences per document. Each article has at least two abstractive summaries written by human annotators, and each summary consists of at most 100 words. 

The SummBank dataset \cite{radev2003summbank} is the best dataset there is at this time for evaluating sentence-ranking algorithms.
It provides benchmarks produced by three human judges. The judges annotated 200 news articles written in English with an average of 20 sentences per document, resulting in, for each article, three sets of sentence rankings, one by each judge.
Ranking scores of sentences can be derived from their rankings. 
In addition, SummBank also provides, for each article, a set of combined sentence ranking of all judges, where the combined ranking of each sentence is the average ranking scores of all three judges. 

\subsection{The ROUGE measures}

ROUGE \cite{lin2004rouge} is a widely used metric to evaluate accuracy of summaries.  
ROUGE-n is an n-gram recall between the automatic summary and a set of references, where 
ROUGE-SU4 evaluates an algorithm-generated summary using skip-bigram and unigram co-occurrence statistics, allowing at most four intervening unigrams when forming skip-bigrams.

An ultimate objective for any machine ranking method would be to achieve the highest possible ROUGE measures against rankings of all judges, which are served as references.  In particular, in the SummBank dataset,  if the corresponding mean ROUGE scores of a machine ranking and
the combined ranking of all judges against the three reference rankings are
comparable, then the machine ranking is deemed as good as the combined wisdom of all three human judges.

\subsection{Comparisons on DUC-02}

On the DUC-02 dataset, 
SWR, SPR, and SSR extract, respectively, sentences of the highest ranks with a total length bounded by 100 words. The results under ROUGE-1 (R-1), ROUGE-2 (R-2), and ROUGE-SU4 (R-SU4) against the DUC-02 benchmarks are shown in Table \ref{tab:1}. The corresponding scores published in $CP_3$ \cite{parveen2016generating}  and a number of major algorithms before it are also presented.
In the table, the highest scores are shown in boldface,
where UA means that the value is unavailable in the corresponding publications.
Note that $CP_3$ outperforms all algorithms before it under these ROUGE measures.

\begin{center}
\captionof{table}{Comparison results (\%) on DUC-02}
\label{tab:1}
\begin{tabular}{l|c|c|c}
\hline \bf Methods & \bf ~R-1~ & \bf ~R-2~ & \bf ~R-SU4 \\ \hline
SSR 				& \bf 49.3 	& \bf 25.1 	& \bf 26.5 \\
SPR 				& 49.2 			& 25.0 		& 26.3 \\
SWR 			& 49.2 			& 24.7 		& 26.1 \\
$CP_3$				& 49.0			& 24.7		& 25.8 \\
CNN-W2V 	& 48.6			& 22.0		& UA	\\
$E_{Coh.}$	& 48.5			& 23.0		& 25.3	\\
URank			& 48.5			& 21.5		& UA	\\
$T_{Coh.}$	& 48.1			& 24.3		& 24.2	\\
TextRank		& 47.1 			& 19.5		& 21.7	\\
ULink ($k$=10)~ & 47.1			& 20.1		& UA	\\ \hline
\end{tabular}
\end{center}
\smallskip
The following results can be seen against the DUC-02 benchmarks:
\begin{enumerate} 
\item SSR outperforms SPR under every measure.
\item SPR 
outperforms SWR under ROUGE-2 and ROUGE-SU4, and has the same ROUGE-1 score as SWR.

\item
SWR outperforms $CP_3$ under ROUGE-1 and ROUGE-SU4, and has the same
ROUGE-2 score as $CP_3$.
\end{enumerate}

\subsection{Comparisons on SummBank}

Comparisons are made with each individual judge's ranking and the combined ranking of sentences of all judges. The combined ranking of sentences on a given document is obtained as follows: First derive individual judges' ranking scores for each sentence contained in the document, then average individual ranking scores as the combined
ranking score of the sentence.
\begin{itemize}
\item To compare with each judge's ranking of sentences, for Judge $i$ ($i=1,2,3$), 
the evaluation uses the other two judges' rankings of sentences as references.
\item To compare with the combined ranking of sentences by all judges, the sentence rankings of all individual judges are used as references.  
\end{itemize}

Table \ref{tab:2} depicts the ROUGE-1, ROUGE-2, and ROUGE-SU4 scores of different methods
on selections of top 5\% of sentences.

\begin{center}
\captionof{table}{ROUGE (\%) comparisons results on SummBank with the 5\% constraint on sentence selections 
}
\label{tab:2}
\begin{tabular}{l|c|c|c}
\hline \bf Methods &\bf R-1  &\bf R-2  &\bf R-SU4 \\\hline
\bf Judge 1	&38.42		&28.20		&26.61 \\
SSR				&\bf 61.34		&\bf 51.61		&\bf 50.54 \\
SPR				&60.51		&51.20		&50.12 \\
SWR				&59.42		&50.69		&49.89 \\
TextRank		&49.52		&38.58		&37.76 \\
\hline
\bf Judge 2	&30.79		&19.95		&19.38 \\
SSR				& \bf 50.03	&\bf 39.52	 &\bf 38.31 \\
SPR				&48.81		&38.09		&37.62 \\
SWR				&47.51		&37.68		&36.18 \\
TextRank		&43.10		&32.66		&31.42 \\
\hline
\bf Judge 3	&35.74		&25.86		&24.66 \\
SSR				&\bf 54.22	&\bf 44.39	 &\bf 43.91 \\
SPR				&53.83	 	&43.44		&42.87 \\
SWR				&53.01	 	&43.38		&42.11 \\
TextRank	&45.81	 &34.32	&33.08 \\
\hline
\bf Combined	&51.60			&\bf 43.50		&\bf 41.50 \\
SSR					&\bf 51.66		&43.32		&41.32 \\
SPR					&51.14			&42.36		&40.53 \\
SWR					&50.92	 		&41.6		&40.31 \\
TextRank			&44.66	 		&33.63		&32.56 \\
\hline
\end{tabular}

\end{center}
\smallskip

\end{multicols}

\begin{figure}[H]
\centering
\begin{tabular}{cc}
 \includegraphics[width=0.5\textwidth]{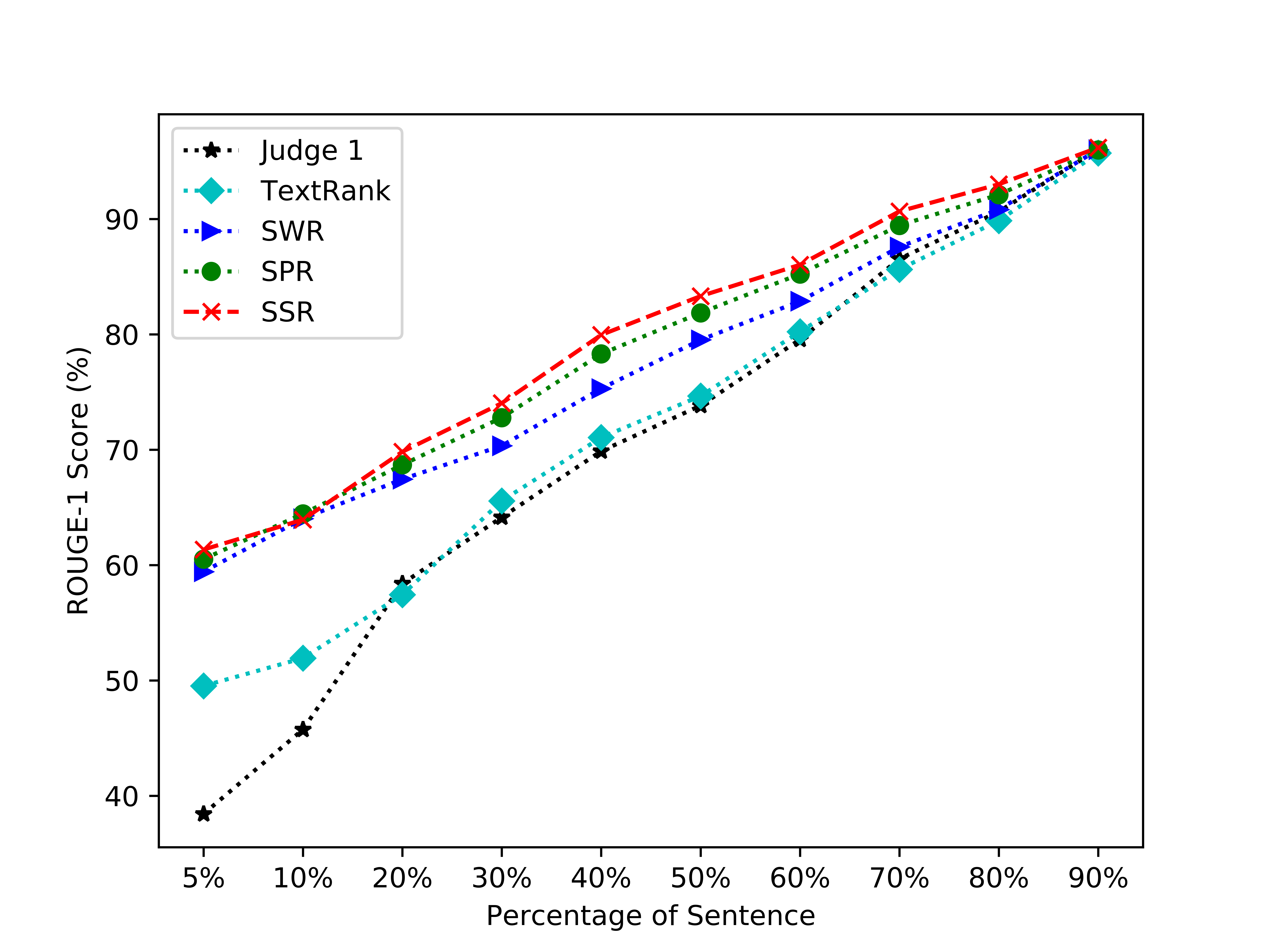}
&
\includegraphics[width=0.5\textwidth]{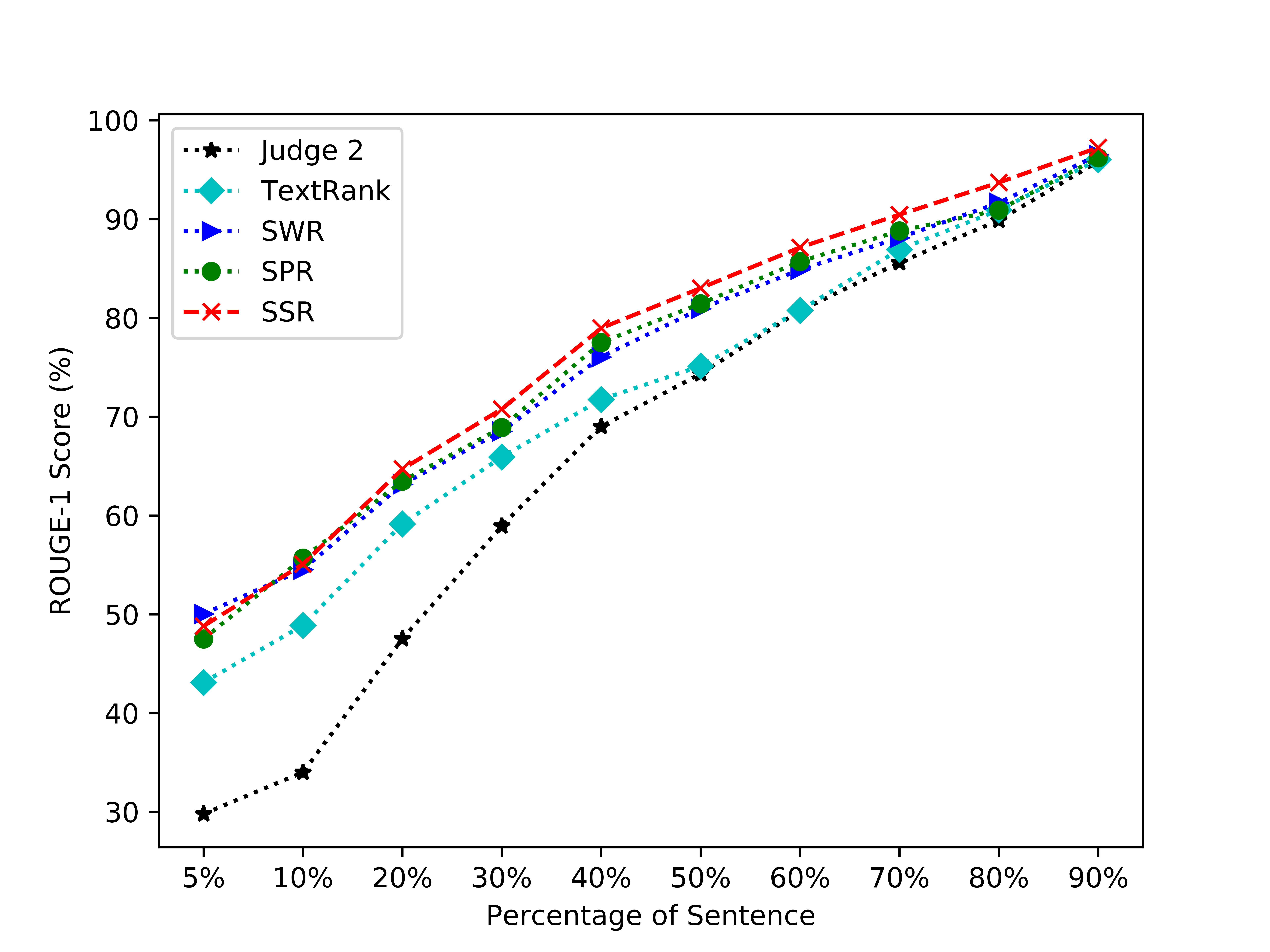} \\
(a) Judge 1 & (b) Judge 2 \\
\includegraphics[width=0.5\textwidth]{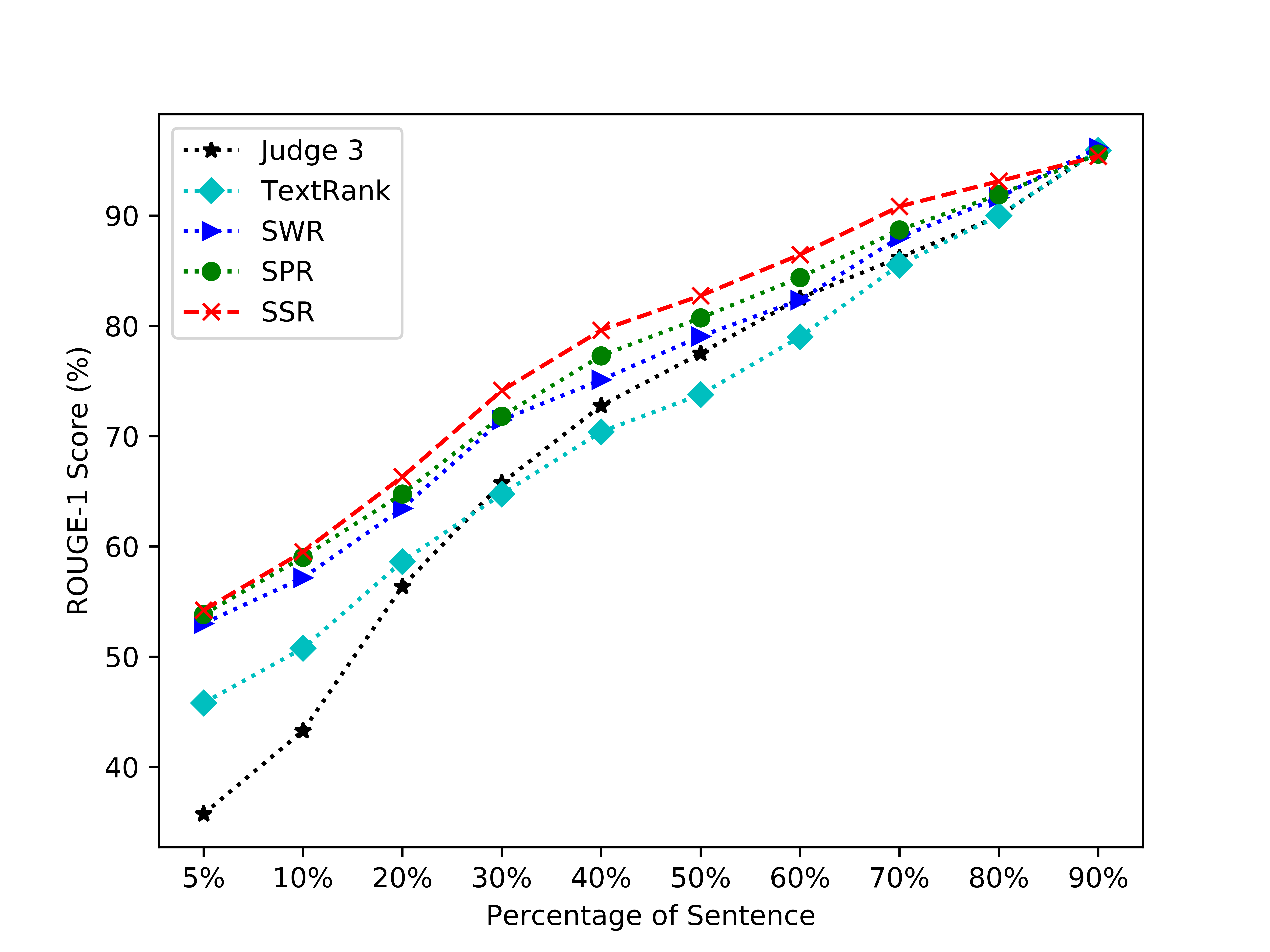}
&
\includegraphics[width=0.5\textwidth]{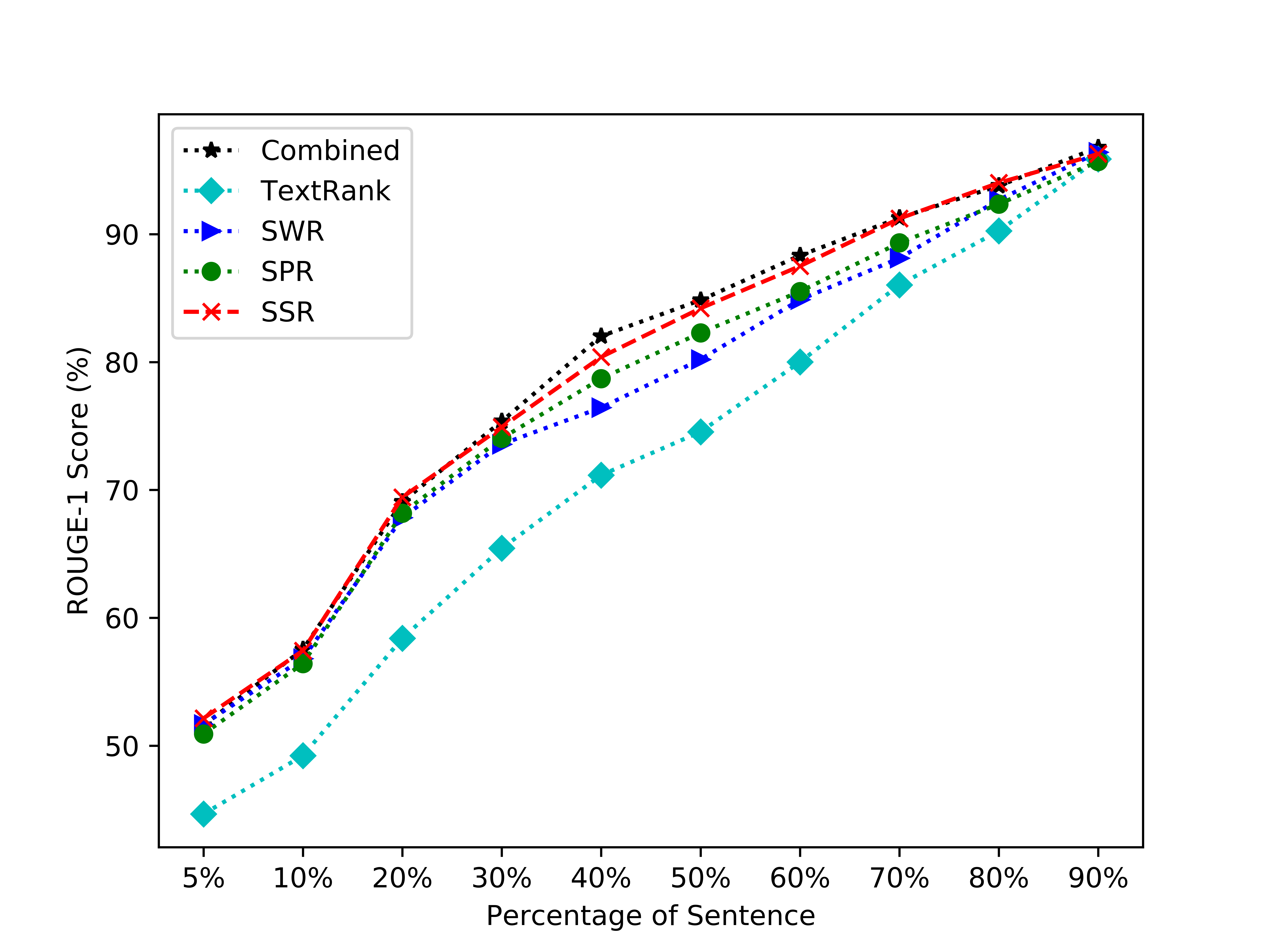} \\
(a) Judge 3 & (b) Combined ranking of all judges \\
\end{tabular}
\caption{ROUGE-1 (\%) comparisons of individual judges and combined ranking
with TextRank, SWR, SPR, and SSR over the SummBank benchmarks: (a) Judge 1 against Judges 2 and 3; (b) Judge 2 against Judges 1 and 3; (c) Judge 3 against Judges 1 and 3; (d)
Combined ranking against all judges}
\label{fig:1}
\end{figure}

\begin{multicols}{2}
The following results are evident:

\begin{enumerate}
\item Under all categories, SSR outperforms each judge by a significant margin and also outperforms SPR, which outperforms SWR, and SWR significantly outperforms TextRank.
\item SSR slightly outperforms the combined ranking of all judges under ROUGE-1, and is slightly below but very close to the combined ranking under ROUGE-2 and ROUGE-SU4, with
the percentage differences being, respectively,  0.029\%,	0.104\%, and 0.109\%.
Moreover, SPR is slightly below SSR,
SWR is slightly below
SPR, and TextRank is substantially below SWR.
\end{enumerate}

A full range of comparisons under ROUGE-1 with individual judges and the combined ranking of 
all judges are given at Fig. \ref{fig:1}, 
where the percentage indicates that a portion of sentences are selected according to their ranks 
by the underlying methods. Thus, when 90\% or more sentences are selected,
all methods are comparable.

{
To demonstrate the robustness of an algorithm, 
it is customary to also compare ROUGE-2 and ROUGE-SU4 scores against the corresponding
\parfillskip=0pt\par}

\end{multicols}

\begin{sidewaystable}[ph!]
\vspace*{1in}
\caption{Full-range comparisons of SSR, SPR, SWR, and TextRank with individual judges and the combined ranking of all judges over the SummBank benchmarks, where shaded
numbers in a group of comparisons for an individual judge are the highest scores under the corresponding measures, while in the group of comparisons for combined ranking, the shaded numbers are the highest and the second highest scores}
\includegraphics[width=\textwidth]{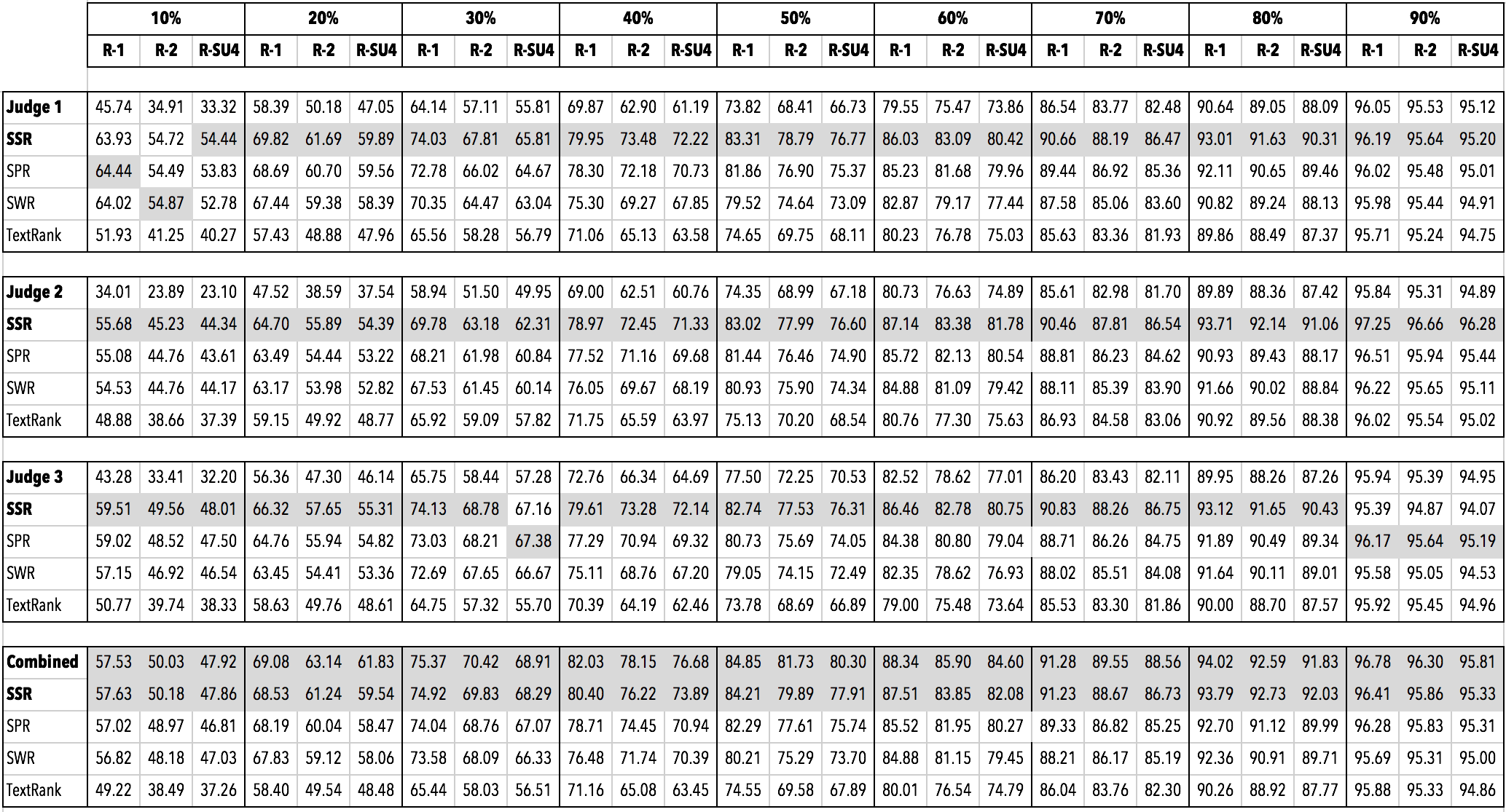}
\label{table:full-range-comparisons}
\end{sidewaystable}

\begin{multicols}{2}
\noindent references.
An algorithm is robust if it compares consistently against the references under different ROUGE measures.
Full-range comparisons under the ROUGE-2 and ROUGE-SU4 measures 
(see Table \ref{table:full-range-comparisons})
show similar trends to
those under ROUGE-1, indicating that SSR, SPR, and SWR are robust. 
Table \ref{table:full-range-comparisons} shows the comparison results 
from 10\% top-ranked sentences to 90\%, with an increment of 10\% each time.

It follows from Table  \ref{table:full-range-comparisons} that, under the ROUGE-1, ROUGE-2, and ROUGE-SU4 measures,  comparison results similar to those on
the 5\% top-ranked sentences discussed on Table \ref{tab:2} hold true on the entire spectrum.
More specifically,

\begin{enumerate} 

\item 
SSR is better than SPR and
 is comparable with the combined ranking of all judges 
at all percentage levels with slightly smaller scores.

\smallskip
\item SPR is better than SWR and narrows the gap between SWR and the combined ranking. 

\smallskip
\item SWR is significantly better than TextRank.

\begin{enumerate}
\item SWR is substantially better than each individual judge's ranking. 

\item SWR is compatible on top ranked sentences (up to 30\%), but incurs a moderate gap on lower ranked sentences between 30\% and 90\%.
\end{enumerate}

\item TextRank is substantially worse than
the combined ranking of all judges. On the other hand, TextRank is better than individual judges on sentences of higher ranks but worse 
on sentences of lower ranks. 
\begin{enumerate}
\item Comparison with Judge 1: TextRank is better on top ranked sentences (up to 20\%) and
worse on the rest of the sentences. 

\item Comparison with Judge 2: TextRank is much better on top ranked sentences (up to 30\%) and slightly better on the rest of the sentences.

\item Comparison with Judge 3: TextRank is better or comparable on top ranked sentences (up to 30\%), worse on lower ranked sentences from 30\% to 70\%, and comparable on the rest of the sentences.
\end{enumerate}
\end{enumerate}

TextRank is based only on co-occurrences of words for computing sentence scores, 
considering neither semantic information, nor subtopic diversity, nor structure of the underlying document.
Incorporating these three features 
is expected to significantly improve 
the accuracy of sentence ranking, and the experiment results confirm that it is true.
While incorporating semantics of words is better than without them,
incorporating semantics of phrases and words is better than just using semantics of words,
and adding semantics of sentences is even better.

It will be shown next how word semantics,
article structure, Softplus function adjustment, and subtopic clustering
each contribute to the improvement of sentence ranking.

\subsection{Significance of each feature}
\label{sec:feature}

It is interesting to understand how each of the features of semantic edges, Softplus function adjustment, ASB PageRank, and subtopic clustering actually contributes to the improvement of sentence rankings. To answer this question it suffices to evaluate the basic model SWR by removing a feature from it one at a time. Let SWR\_NSE, SWR\_NAS, SWR\_NSC, and SWR\_NSP denote, respectively, the variant of SWR without semantic edges, article-structure information, subtopic clustering, and Softplus adjustment.

Table \ref{tab:3} is the results obtained from evaluations over SummBank on selections of 10\%, 40\%, and 70\% of sentences. TextRank is included as a baseline.  The numbers in bold are the most severe drops, indicating that the corresponding features are the most
critical.  

\end{multicols}

\begin{table}[h]
\begin{center}
\caption{ROUGE (\%) comparison with different features removed}
\label{tab:3}
\begin{tabular}{l|c|c|c|c|c|c|c|c|c}
\hline
\bf \multirow{2}{*}{Methods} & \bf R-1 & \bf R-2 & \bf R-SU4  & \bf R-1 & \bf R-2 & \bf R-SU4 & \bf R-1 & \bf R-2 & \bf R-SU4\\ \cline{2-10}
& \multicolumn{3}{c|}{10\%} & \multicolumn{3}{c|}{40\%} & \multicolumn{3}{c}{70\%}\\ \hline
SWR 	&	56.82	&	48.18	&	47.03   &76.48	&	71.74	&	70.39  &88.21	&	86.17	&	85.19\\
SWR\_NSE 	&	55.02	&	46.08	&	45.15  & \bf71.98	&	\bf66.76	&	\bf65.42	 &\bf85.94	&	\bf83.18	&	\bf81.83\\
SWR\_NAS 	&	\bf51.74	&	\bf40.87	&	\bf40.13  &73.52&	68.26	&	67.12  &87.82	&	84.95	&	84.12\\
SWR\_NSC 	&	56.36	&	47.34	&	46.44 & 75.68	&	70.35	&	69.03 & 87.54	&	84.49	&	83.54\\
SWR\_NSP	&	56.77	&	48.12	&	46.97 &	76.39	&	71.59	&	70.25 & 88.14	&	86.04	&	85.07\\
TextRank 	&	49.22	&	36.07	&	35.29 & 71.16	&	65.47	&	63.94 & 86.04	&	83.52	&	82.29\\
\hline
\end{tabular}
\end{center}
\end{table}

\begin{multicols}{2}
The following results are drawn:
\begin{enumerate}
\item With each of the features removed, the corresponding ROUGE scores drop, indicating that each feature has contributed to the improvement. 
\item when the percentage of selecting sentences is smaller, removing article structure results in much larger drops, indicating that article structures are more significant on top-ranked sentences. 

\item When the percentage becomes larger, semantic edges would become increasingly more critical. 

\item While the Softplus function adjustment does improve ROUGE scores, it is not as significant as the other features.

\item Each SWR variant outperforms TextRank under all ROUGE measures, except that at 
the 70\% level, when semantic edges are removed, SWR\_NSE is lightly below TextRank.
\end{enumerate}

Note that it is difficult to capture the underlying meanings of polysemous words using word-embedding representations, for one word-embedding representation cannot reflect different meanings. To train different word-embedding representations for a polysemous word so that each representation corresponds to only one meaning,
it needs a corpus of documents containing the word only
in one meaning. This is a formidable task because a document may contain multiple 
polysemous words, and each of these words may actually appear with one meaning here and a different meaning there in the same document. 
Even if multiple word-embedding representations could be trained for a polysemous word, it would still be challenging to determine which embedding representation should be used for a particular occurrence of the word. Using co-occurrences
of words and phrases can help recoup the underlying meaning of a polysemous word
lost in its word-embedding representation.

\section{Conclusions and final remarks} \label{sec:9}

SSR is an efficient and accurate scheme for ranking sentences of a given document. In particular, an implementation of SSR presented in this paper runs in quadratic time, and outperforms,
 on the SummBank benchmarks,  each individual human judge's ranking under standard ROUGE measures and
compares well with the combined ranking of all judges. Moreover, extracting sentences of the highest ranks with an appropriate number of sentences as an extractive summary achieves the state-of-the-art results over the DUC-02 benchmarks under standard ROUGE measures.

SSR is an unsupervised scheme that does not rely on language-specific features or deep linguistic computations, and so it is readily adaptable across languages. What it needs is a reasonable corpus of digital documents available in the adapted language (such as Wikipedia dumps) for extracting phrases and training word and phrase embedding representations. However, the similarity threshold values presented in this paper
for constructing a semantic word graph and a semantic phrase graph for a given document, while
appropriate for the English language and not sensitive to a small change, may need to be adjusted for other languages.
Better threshold values can be determined using a small amount of labeled data. 
When no labeled data is available, an expected number of synonyms may be used to determine appropriate thresholds \cite{rekabsaz2017exploration}.

It would be interesting to figure out how much weight should be given to the word-embedding similarity in a semantic-word graph and a semantic-phrase graph to better reflect the underlying meaning of a polysemous word. It would also be interesting to seek if there are more effective, unsupervised methods,
incorporating nearby monosemous words
in the context, to 
help bring out the underlying meaning of a polysemous word. The accuracy of the word-embedding-based
WMD semantic measure of two sentences used in a semantic-sentence graph also suffers from polysemous words contained in them. A better way to handle polysemous words would be needed for improvement.

Not reported in this paper, location-score functions for sentences can be further improved to better reflect the structure of the underlying article in the following four categories: narration, argumentation, research, and news. Improved location-score functions have been implemented
at \url{http://dooyeed.com}. 

The following directions may be explored for further improvement of accuracy, in addition to what has been mentioned in the earlier sections:
\begin{enumerate}
\item Investigate other embedding methods such as ELMo \cite{peters2018deep}, LEAR
\cite{vulic2017specialising}, Poincar\'{e} embeddings \cite{nickel2017poincare}, hierarchical embedding \cite{ai2017learning}, and spherical embedding \cite{batmanghelich2016nonparametric}.
\item Investigate unsupervised sentence representations such as skip-thought vectors \cite{kiros2015skip} and BERT \cite{devlin2018bert}.
\item Fine-tune location scoring for better representing article structures.
\item Explore new approximation algorithm for the NP-hard multi-objective
knapsack problem. For example, instead of using round-robin approximation that 
treats each cluster equally likely, a weighted round-robin strategy that treats
each cluster according to the subtopic distribution in the given document may
help improve accuracy.
\end{enumerate}

Finally, readers should keep in mind that employing new methods, while possibly improving accuracy, may also degrade efficiency. Whether a trade-off is acceptable depends on the underlying applications.

\section*{Acknowledgments}
This work was supported in part by Eola Solutions, Inc.
The authors thank Wenjing Yang, Yicheng Sun, and Changfeng Yu of UMass Lowell for writing an API to run the AutoPhrase code available at Github to carry out SSR evaluations, and to Liqun (Catherine) Shao of Microsoft for an interesting discussion on Softplus function adjustment. 
They are grateful to Cheng Zhang at UMass Lowell for implementing dooyeed.com that uses SSR to rank sentences for a given document.
The second author would also like to thank Jiawei Han of the University of Illinois at Urbana-Champaign and Jesse Wang of the University of Rochester for inspiring conversations.

\bibliographystyle{vancouver}
\bibliography{SSR}

\end{multicols}

\end{document}